\documentclass[twoside]{article}

\usepackage{qic,epsfig}
\usepackage{amsmath}
\usepackage{cite}

\begin{document}
\setlength{\textheight}{8.0truein}

\runninghead{Blind Reconciliation}
            {J. Martinez-Mateo, D. Elkouss and V. Martin}

\normalsize\textlineskip
\thispagestyle{empty}
\setcounter{page}{1}

\copyrightheading{0}{0}{2003}{000--000}

\vspace*{0.88truein}

\alphfootnote

\fpage{1}

\centerline{\bf BLIND RECONCILIATION}
\vspace*{0.37truein}
\centerline{\footnotesize JESUS MARTINEZ-MATEO, DAVID ELKOUSS, VICENTE MARTIN\footnote{vicente@fi.upm.es}}
\vspace*{0.015truein}
\centerline{\footnotesize\it Research group on Quantum Information and Computation}
\baselineskip=10pt
\centerline{\footnotesize\it Facultad de Inform\'{a}tica, Universidad Polit\'ecnica de Madrid}
\baselineskip=10pt
\centerline{\footnotesize\it Campus de Montegancedo, 28660 Boadilla del Monte, Madrid, Spain}
\vspace*{0.225truein}
\publisher{(received date)}{(revised date)}

\vspace*{0.21truein}
\abstracts{Information reconciliation is a crucial procedure in the classical post-processing of quantum key distribution (QKD). Poor reconciliation efficiency, revealing more information than strictly needed, may compromise the maximum attainable distance, while poor performance of the algorithm limits the practical throughput in a QKD device. Historically, reconciliation has been mainly done using close to minimal information disclosure but heavily interactive procedures, like \textit{Cascade}, or using less efficient but also less interactive ---just one message is exchanged--- procedures, like the ones based in low-density parity-check (LDPC) codes. The price to pay in the LDPC case is that good efficiency is only attained for very long codes and in a very narrow range centered around the quantum bit error rate (QBER) that the code was designed to reconcile, thus forcing to have several codes if a broad range of QBER needs to be catered for. Real world implementations of these methods are thus very demanding, either on computational or communication resources or both, to the extent that the last generation of GHz clocked QKD systems are finding a bottleneck in the classical part. In order to produce compact, high performance and reliable QKD systems it would be highly desirable to remove these problems. Here we analyse the use of short-length LDPC codes in the information reconciliation context using a low interactivity, \textit{blind}, protocol that avoids an a priori error rate estimation.  We demonstrate that $2 \times 10^3$ bits length LDPC codes are suitable for blind reconciliation. Such codes are of high interest in  practice, since they can be used for hardware implementations with very high throughput.}
{}
{}
\vspace*{10pt}

\keywords{Quantum key distribution, information reconciliation, low-density parity-check codes, rate-compatible, interactive reconciliation, short-length codes}
\vspace*{3pt}
\communicate{to be filled by the Editorial}

\vspace*{1pt}\textlineskip

\section{Introduction}

Quantum key distribution (QKD) \cite{Gisin_02} is a process that runs in two phases: a quantum and a classical one. The second one distills a secret key from the quantum signals produced and transmitted in the first. The distillation process begins with an information reconciliation step: from two correlated strings at both ends of the channel, a common string is extracted by publishing some amount of information.

In this context, a set of outputs obtained from measurements of quantum states, must be cooperatively processed at both ends of the transmission channel in order to obtain the common secret key from which all information leakage has been bounded \cite{Gottesman_04}. The set of measurements ---raw key--- have errors due to imperfections in the devices, decoherence, eavesdropping or the limited efficiency of the protocol itself. These data strings must be reconciled in order to have the same string at both ends, a process that by revealing information through a public but noiseless channel removes any discrepancy and is known as \textit{information reconciliation} \cite{Brassard_94}.

In order to further clean the reconciled key from any information leakage, either by the previous procedure or by the exchange of quantum signals, a second step is required. This is known as \textit{privacy amplification} \cite{Bennett_95} and generates the final secret key, shorter but with a known upper bound on the information leakage. Thus, the final secret key length is dependent also on the quality of the information reconciliation step and its behavior regarding security related parameters (e.g. finite size effects).

Information reconciliation in QKD is a problem already addressed by the authors of the original BB84 protocol \cite{Bennett_88, Bennett_92, Brassard_94}. In the pioneering BBBSS protocol \cite{Bennett_92}, the authors propose a reconciliation protocol based in the exchange of a number of parities (syndromes). If any parity-check equation of an exchanged syndrome is not verified, the parties carry out a dichotomic search to find the corresponding error. Note that in each parity-check equation an odd number of errors can be detected, but only one of them can be corrected using a binary search. The procedure works iteratively, shuffling the bits of the key to reconcile and exchanging successive syndromes. Later, in Ref.~\cite{Brassard_94}, the authors realized that each located error produces side information that can be used with a previously exchanged syndrome. The new protocol was called \textit{Cascade} in reference to the iterative or cascading process of identifying errors. Several optimizations were proposed for the BBBSS and \textit{Cascade} protocols \cite{VanDijk_97, Sugimoto_00, Liu_03}. However, all these protocols are highly interactive since they require many communication rounds: the parties have to exchange a large number of messages. Despite its interactivity, \textit{Cascade} continues being one of the most widely used protocols for information reconciliation in QKD, probably due to its simplicity and relatively good efficiency.

Other protocols have been proposed in the literature. For instance, in \textit{Winnow} the authors use Hamming codes for the calculation of separate syndromes instead of a simple parity-check equation \cite{Buttler_03}. However, the efficiency of this protocol is still far from the Shannon limit.

As early as 2003, the Alamos group hinted at the use of parity-checks as in telecommunication systems \cite{Elliott_03}, but the group did not present any result referring to the use of low-density parity-check (LDPC) codes until one year later \cite{Pearson_04, Elliott_05}. This is one of the first applications of the modern coding theory to the information reconciliation problem in QKD.

The objective of this work is to produce an information reconciliation method amenable to practical implementation using modern hardware for embedded systems. We base our approach in the use of rate adaptive low-density parity-check (LDPC) codes. In  Ref.~\cite{Elkouss_11} we propose a reconciliation protocol for large codes ($2 \times 10^5$) in order to obtain the best possible efficiency. These codes have the advantage of minimal interactivity, thus avoiding one of the main disadvantages of \textit{Cascade}. In Ref.~\cite{Martinez_10} we studied the efficiency improvement when relaxing the condition of minimal interactivity, also for large codes. However, neither method is appropriate for a hardware implementation. The purpose of the present paper is to adapt those methods in order to cover a varying error range with high efficiency and, at the same time, make them suitable for a hardware implementation. The limited resources available in embedded hardware make unfeasible the use of a long-length code, thus we focus our interest in the use of short-length LDPC codes. Here we will be using a code length as short as $2 \times 10^3$, two orders of magnitude smaller than in the previous works. This implies that the techniques used in the previous algorithms are no longer optimal. For instance, the random puncturing methods used previously can reduce to zero the distillable secret key rate, thus new schemes had to be devised \cite{Elkouss_11b}. Moreover, we demonstrate in this correspondence that by allowing a limited interactivity, the final efficiency of the reconciliation can improve significantly. In Ref.~\cite{Martinez_10} we studied the limiting case where only one bit changes per step, without regard of minimizing the number of steps. Here we minimize the interactivity, demonstrating that  as little as three messages can approach the efficiency of the maximally interactive case. This makes possible a relatively high throughput process implementable with limited resources. For example, three relatively small FPGA blocks that can work in parallel will suffice. Remarkably, the new protocol works without an a priori estimation of the quantum bit error rate, a fact with interesting implications in finite key analysis: since no extra information is revealed, it does not need to be subtracted from the key.

This correspondence is organised as follows. In Section~\ref{sec:reconciliation} we introduce the information reconciliation problem and its application using low-density parity-check codes. In Section~\ref{sec:rate-adaptive} we review some techniques used for adapting the information rate of a correcting code, and reduce the information disclosed in the reconciliation when using LDPC codes. In Section~\ref{sec:interactive} we present an interactive version of a rate-adaptive protocol (that we call \textit{blind}) that improves the average efficiency. In Section~\ref{sec:results} we show some simulation results of the blind protocol using short-length LDPC codes. Finally, in Section~\ref{sec:conclusions} we present our conclusions.

\section{Information Reconciliation and Channel Coding}
\label{sec:reconciliation}

In this section we consider the problem of information reconciliation from an information theoretic perspective and study some figures of merit relevant to the discussed protocols.

Information reconciliation is the generic name of any method used to ensure that two parties agree on a common string provided they have access to two correlated sequences $X$ and $Y$ \cite{VanAssche_06}. During reconciliation the two parties exchange a set of messages $M$ over a noiseless channel such that at the end of the process they agree on some string function of their strings and the exchanged messages. In our case, the correlated strings are obtained by Alice and Bob after the quantum phase of the QKD protocol has finished. It does not matter whether an actual quantum channel has been used to transmit qubits from Alice to Bob as in a standard prepare and measure protocol or an entangled pairs emitter acts as the source of correlations. In both cases, $X$ and $Y$ can be regarded as correlated random variables and every symbol in $Y$ can be seen as given by transition probability $p_W(y|x)$, or equivalently as if every symbol were the output of a memoryless channel $W$.

Typically, channels are classified in families characterized by some continuous variable, $\epsilon$, selected to parameterise its behaviour. The variable $\epsilon$ is chosen such that increasing values of $\epsilon$ imply a degraded version of the channel \cite{Richardson_01a}. For example, the family of binary symmetric channels (BSC) is parameterised by the error rate and the family of additive white Gaussian noise (AWGN) channels by the noise variance. A channel $W_{\epsilon '}$ is a degraded, or noisier, version of the channel $W_{\epsilon}$ if:

\begin{equation}
\label{eq:degradation}
p_{W_{\epsilon '}}(y'|x) = p_{Q}(y'|y) p_{W_{\epsilon}}(y|x)
\end{equation}

\noindent where $Q$ is some auxiliary channel.

Let the information rate be the proportion of non redundant symbols sent through a channel. A code $\mathcal{C}(n,k)$, defined by a parity check matrix $H$, transforms an string of $k$ symbols in a codeword $c$ of $n$ symbols with $k$ independent symbols and $n-k$ redundant symbols, and in consequence achieves an information rate, $R$, of $k/n$.

This parameterisation allows to study two related concepts: the capacity of a channel, that is the maximum information rate that can be transmitted for a fixed $\epsilon$ and, for a specific error correcting code $\mathcal{C}$, the maximum value $\epsilon_{\textrm{max}}$, i.e. the noisiest channel for which a sender can reliably transmit information with $\mathcal{C}$. The relationship between both answers gives an idea of the efficiency of the code, or in other words, how close is the coding rate of a code to the optimal value.

We can measure, analogously, the efficiency $f$ of an information reconciliation protocol as the relation between the length of the messages exchanged to reconcile the strings, $|M|$, and the theoretical minimum message length. The problem of information reconciliation in secret key agreement is formally equivalent to the problem of source coding with side information \cite{Slepian_73}, or how should $X$ be encoded in order to allow a decoder with access to side information $Y$ to recover $X$. Thus, the minimum message length is given by the conditional entropy $H(X|Y)$, since given the decoder access to side information $Y$ no encoding of $X$ shorter than $H(X|Y)$ allows for reliable decoding \cite{Slepian_73}. We can define the efficiency of an information reconciliation procedure as:

\begin{equation}
\label{eq:efficiency}
f = \frac{|M|}{H(X|Y)}
\end{equation}

\noindent where $f=1$ stands, then, for perfect reconciliation.

Error correcting codes can be used for information reconciliation \cite{Crepeau_95}. In information reconciliation, $Y$ is already a noisy version of $X$ (or viceversa) and the encoder and decoder have access to a noiseless channel. A code $\mathcal{C}$ can be used for information reconciliation using \textit{syndrome} coding \cite{Cover_91}. The syndrome of $\mathbf{x}$, an instance of $X$, is defined as $s(\mathbf{x}) = H \cdot \mathbf{x}$, with length per symbol $1-R$, indicates in which of the cosets of $\mathcal{C}$ is $\mathbf{x}$ a codeword. In other words, a decoder should recover $\mathbf{x}$ with high probability given access to $\mathbf{y}$, an instance of $Y$, and $s(\mathbf{x})$, the syndrome of $\mathbf{x}$, in a code adapted to the correlation between $X$ and $Y$. Then, the efficiency of an information reconciliation method based in syndrome coding is given by:

\begin{equation}
\label{eq:efficiency-source-coding}
f_{\mathcal{C}} = \frac{1 - R}{H(X|Y)}
\end{equation}

\noindent which in the special, but important, case of a binary symmetric channel with error rate $\epsilon$, BSC($\epsilon$), can be written as:

\begin{equation}
\label{eq:efficiency-bsc}
f_{\mathrm{BSC}(\epsilon)} = \frac{1-R}{h(\epsilon)}
\end{equation}

\noindent where $h(\epsilon)$ is the binary Shannon entropy.

\begin{figure} [htbp]
\centerline{\epsfig{file=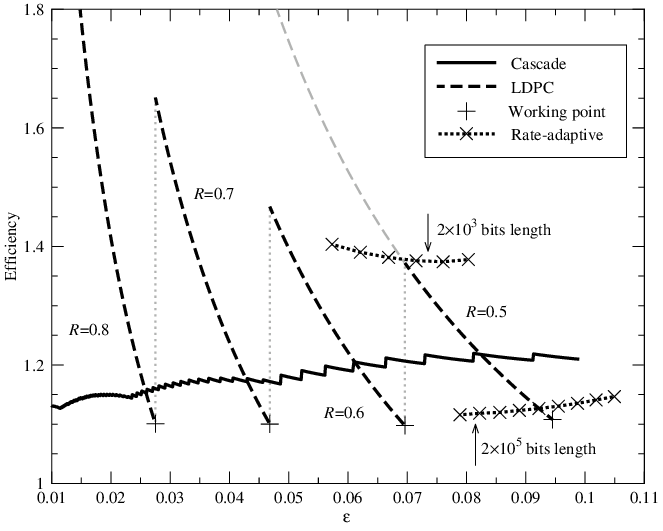, width=0.8\textwidth}} 
\vspace*{13pt}
\fcaption{Reconciliation efficiency using \textit{Cascade}~\cite{Brassard_94} and with LDPC codes for a $2 \times 10^5$ bits key length as a function of the channel error rate, $\epsilon$. The short dashed curve marked as LDPC shows a set of codes where the best efficiency code is chosen for every value of $\epsilon$. When $\epsilon$ is just below or at the ``working'' point, all the redundancy is used, producing high efficiency codes. As we move away from this point using the same code, the efficiency moves further from the optimal value. To put these data in context, we show with dotted lines the efficiency of a rate adaptive solution for R=0.5 using both, a short code of $2 \times 10^3$ bits length and a code of length $2 \times 10^5$. The proportion of symbols used to modulate the rate was set to $10\%$. Note that the length of the codes is two orders of magnitude smaller, which allows for hardware implementation. Since the rate modulation is performed by puncturing and shortening (simultaneously) the curve for the rate adaptive approach is centered on the curve for the non adaptive reconciliation with identical coding rate and length.}
\label{fig:saw}
\end{figure}

The behaviour of the reconciliation efficiency using a code $\mathcal{C}$ as a function of the characteristic parameter, here the error rate $\epsilon$, is shown in Fig.~\ref{fig:saw}. The efficiency decreases in the range $\epsilon \in [0,\epsilon_{\textrm{max}}]$. Since the redundancy is fixed for a $\epsilon$ range, it is excessive and far from optimal for good channels, i.e. low values of $\epsilon$; as the light gray line shows for a reconciliation method based in a single LDPC code of $2\times 10^5$ bits length. We can improve the reconciliation efficiency using a set of LDPC codes, as in the line marked LDPC, where we have chosen the code with the best efficiency for every value of $\epsilon$, we observe a characteristic saw behaviour: the efficiency is good for $\epsilon$ values just below the $\epsilon_{\textrm{max}}$ of every code and it degrades till the next code is used. This forces the use of many codes in order to cover a broad range of $\epsilon$ with good efficiency, not a very practical proposition. These two solutions based in LDPC codes are compared to a rate-compatible solution developed in the next section. The rate compatible solution has been calculated using short-length LDPC codes ($n = 2 \times 10^3$) which are suited for hardware implementations. There is a tradeoff between efficiency and code length, one of the aims of this work is to increase the reconciliation efficiency using short-length codes. In the figure, the solid line depicts the efficiency of \textit{Cascade} \cite{Brassard_94}, the \textit{de facto} standard for information reconciliation in QKD. The \textit{Cascade} protocol, though highly interactive, has the main advantage of being easy to implement and efficient enough for reconciling short strings. However, its interactivity can easily become a bottleneck in practical QKD systems, specially in those working at high speed or in a high QBER regime.

\section{Rate-Adaptive Reconciliation}
\label{sec:rate-adaptive}

In the previous section we described a solution that allows to cover a $\epsilon$ range with several LDPC codes. However, even if a solution based in several LDPC codes is more efficient than a solution with just one code, it is still impractical. On one hand it forces Alice and Bob to store a set of codes, on the other it relies on precise estimations of $\epsilon$: imprecisions in its estimation could lead to use a code unable to reconcile the strings. It would be highly desirable to use a single code able to adapt to different rates. In this section we review two techniques that allow to adapt the coding rate: \textit{puncturing} and \textit{shortening}.

\subsection{Puncturing}
\label{subsec:punct}

Puncturing modulates the rate of a previously constructed code, $\mathcal{C}(n,k)$, by deleting a set of $p$ symbols from the codewords, converting it into a $\mathcal{C}(n-p,k)$ code (see Refs.~\cite{Ha_04, Pishro-Nik_07}). The rate is then increased to:

\begin{equation}
R (p) = \frac{k}{n-p} = \frac{R_0}{1-\pi}
\label{eq:rate-puncturing}
\end{equation}

\noindent where $R_0=k/n$ is the rate of the original code and $\pi = p/n$ is the fraction of punctured symbols.

Syndrome coding can then be used to adapt the code rate in the following way. Let $\mathcal{C}(n,k)$ be a code that can correct noise up to $\epsilon_{\textrm{max}}$ for some channel family and let $X$ and $Y$ be two $m$ length strings, with $m=n-p$, correlated as if they were the input and output of a channel characterized by $\epsilon < \epsilon_{\textrm{max}}$. The encoder can send the syndrome in $\mathcal{C}$ of a word $\widehat{X}$ constructed by embedding $X$ in a string of size $n$ and filling the other $p$ positions with random bits. If the new coding rate, $R(p)=R_0/(1-p)$ is adapted to $\epsilon$ the decoder should recover $\widehat{X}$ with high probability. 

We can think of a reconciliation protocol based only in punctured codes: the parties would agree on an acceptable frame error rate (FER) and, depending on their estimation of the error rate, they would choose the value of $p$. If we consider the behaviour of FER as a function of $\epsilon$ for a set of fixed $p$ values, as depicted in Fig.~\ref{fig:fer-puncturing}, this procedure can be regarded as moving along the horizontal axis from one code to the next. However, this way of proceeding has the shortcoming that if the channel is time varying (i.e. $\epsilon$ varies over time), the length of $X$ and $Y$ also varies to accommodate the different values of $p$ needed to adapt the coding rate. We could think of scenarios where $m \gg n$ and $X$ and $Y$ can be divided in packets of length $n-p$ but this clearly does not apply to many situations.

\begin{figure} [htbp]
\begin{center}
\epsfig{file=results/puncturing/rho2000-1-pat-10.ps, width=0.7\textwidth} 
\end{center}
\vspace*{13pt}
\fcaption{Frame error rate over the BSC for a binary LDPC code of $2 \times 10^3$ bits length and rate $R_0=1/2$ as a function of the error rate. Several curves have been simulated for different proportions of punctured symbols. Due to the short code length, the distribution of punctured symbols has been intentionally chosen according to an optimized pattern \cite{Ha_06}. Each point in the graph was calculated using as many codewords as needed till the FER was stable. Depending on the specific FER, this was typically 
between $10^3 /\mathrm{FER}$ and $10^2 /\mathrm{FER}$. }
\label{fig:fer-puncturing}
\end{figure}

\subsection{Shortening}
\label{subsec:short}

Puncturing increases the rate by reducing redundancy. The opposite is achieved through shortening: by increasing the redundancy, the information rate is reduced. This is done by fixing the value of a set of $s$ symbols from the codewords in positions known to encoder and decoder. Shortening, then, converts a $\mathcal{C}(n,k)$ code in a $\mathcal{C}(n-s,k-s)$ one \cite{Tian_05}. The result of simultaneously puncturing $p$ symbols and shortening $s$ symbols in the original code is thus a $\mathcal{C}(n-p-s,k-s)$ code with rate:

\begin{equation}
R = \frac{k-s}{n-p-s} = \frac{R_0 - \sigma}{1 - \pi - \sigma}
\label{eq:rate-compatible}
\end{equation}

\noindent where $\sigma = s/n$ is the proportion of shortened symbols.

Typically only puncturing or shortening are used to adapt the rate of a code. However, when using syndrome coding over time varying channels, using just one of the two has the drawback that modifying the value of $p$ or $s$ implies modifying also the length of the reconciled strings with every code use. The combined application of both techniques allows to fix the size of the strings to reconcile and overcome this problem. In this case, a modulation parameter $d=p+s$ can be set, thus fixing the lengths of $X$ and $Y$ to $n-d$ while allowing to modify $p$ and $s$ in order to adapt to different values of $\epsilon$. Fig.~\ref{fig:fer-shortening} shows the performance of an error correcting code, again depicted as the FER versus the error rate of a BSC using both techniques simultaneously. A short-length LDPC code of $2 \times 10^3$ bits length was used. 

\begin{figure} [htbp]
\centerline{\epsfig{file=results/shortening/rho2000-1-pat-10.ps, width=0.7\textwidth}} 
\vspace*{13pt}
\fcaption{FER over the BSC for a binary LDPC code of length $2 \times 10^3$ bits and rate $R_0=1/2$. Several curves have been simulated for different proportions of punctured and shortened symbols, with $d=200$. The distribution of punctured symbols has been chosen according to a pattern previously estimated as proposed in Ref.~\cite{Ha_06}.}
\label{fig:fer-shortening}
\end{figure}

If we call $\delta$ the proportion of punctured and shortened symbols, $\delta = d/n = \pi + \sigma$, a $\delta$-\textit{modulated} rate-adaptive code covers the range of rates $[R_{\textrm{min}}, R_{\textrm{max}}]$ defined by:

\begin{equation}
R_{\textrm{min}} = \frac{R_0 - \delta}{1 - \delta} \leq R \leq \frac{R_0}{1 - \delta} = R_{\textrm{max}}
\label{eq:rate-range}
\end{equation}

There is a tradeoff between the covered error range, increasing with $\delta$, and the efficiency of the procedure, decreasing with higher $\delta$ values \cite{Elkouss_11}. 

The efficiency of a rate-adaptive protocol that systematically applies puncturing and shortening, one bit at a time, up to a pre-established $\delta$ is depicted in Fig.~\ref{fig:saw}, marked as \textit{Rate-adaptive}. Note how the saw tooth behaviour is eliminated. Note also that, since the codes used are very short ($n = 2\times 10^3$), the efficiency is worse than that of \textit{Cascade}.

\section{Blind Reconciliation}
\label{sec:interactive}

In the rate-adaptive algorithm just outlined, the proportion, $\delta$, of punctured plus shortened symbols is held constant. This proportion is calculated after an error rate (channel parameter) estimation. The only classical communication that is needed among Alice and Bob is one message from Alice to send the syndrome and the shortened information bits. This makes for a close to minimal interactivity protocol that is also highly efficient. Now, if we relax the interactivity condition and allow for a limited amount of communications, the panorama changes significantly. 

Let us start by assuming again a value for $\delta$ covering the range of rates $[R_{\textrm{min}},R_{\textrm{max}}]$ with the code with $R_{\textrm{min}}$ able to correct words transmitted through the noisiest channel expected. 

In a first message, Alice can include only the syndrome and no shortened bits, i.e. all the $d$ symbols that can be either punctured or shortened, are punctured ($\pi = \delta$). If we look at Fig.~\ref{fig:fer-shortening}, where we plot the behavior of FER as a function of $\epsilon$ using different proportions of punctured and shortened symbols, we can see that we are trying to correct errors with the code with the highest FER and highest rate, which is the one with $p=200$.

If the reconciliation fails, no other information than the syndrome has been leaked, since punctured symbols do not disclose information. Alice can then reveal a set of the values of the previously punctured symbols. In this way the rate of the code is reduced, but the decoding success probability is increased. Returning to Fig.~\ref{fig:fer-shortening}, this is like moving along the dotted vertical line and changing the code with $p=200$, $s=0$ ($\mathcal{C}(2000-200, 1000)$) by the code with $p=160$, $s=40$ ($\mathcal{C}(2000-200, 1000-40)$) and using it to correct the same string. Only the previously punctured but now shortened symbols reveal extra information. The protocol runs on the same string by revealing more information on the values of previously punctured symbols till success is achieved (or all the symbols were shortened without syndrome matching and it fails), effectively by using at each iteration codes with lower rate and FER. 

\begin{figure} [htbp]
\centerline{\epsfig{file=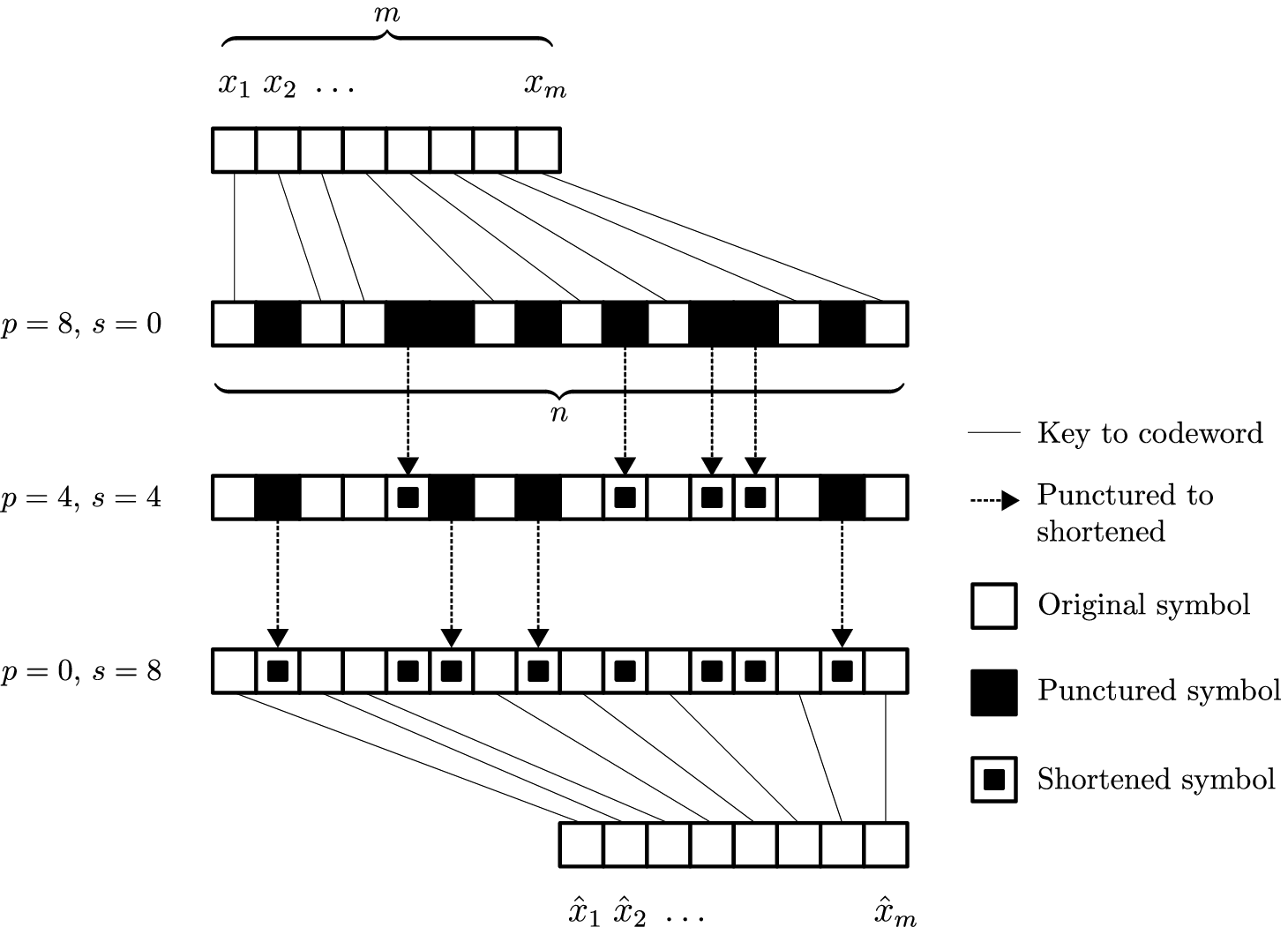, width=0.8\textwidth}} 
\vspace*{13pt}
\fcaption{Interactive protocol (blind), as described in the text.}
\label{fig:interactive-protocol}
\end{figure}

In Fig.~\ref{fig:interactive-protocol} we illustrate two iterations of the protocol in use to reconcile a string of length $m=8$ using $d=8$ extra symbols. It is also assumed that in every iteration $\Delta=4$ symbols can be changed from punctured to shortened. In the first step, the $m$ symbols are incremented with the $d=8$ punctured ones to a total length of $n=m+d=16$. At this point, the syndrome is calculated and the value sent to Bob. It is assumed that there is no syndrome match, hence the next iteration in which $\Delta=4$ of the previously punctured symbols change to shortened. This information is sent to Bob. Again, a no match is assumed and the protocol proceeds to its second iteration, where another 4 symbols are revealed changing from punctured to shortened. Here the protocol ends, no matter whether there is a syndrome match or not, since all the punctured symbols have changed to shortened. If there is a syndrome match, then there is guarantee that the string $(x_1, x_2, \ldots, x_m)$ in the emitter side and $(\hat{x}_1, \hat{x}_2, \ldots, \hat{x}_m)$ in Bob's side are the same. Otherwise, the protocol fails for this string.

This whole procedure is done using the same base code and without needing an estimate for $\epsilon$, hence the \textit{blind} name. Only a rough estimate of the channel parameter is needed to design the base code. Note that this protocol requires some interactivity since, at each iteration in which there is no syndrome matching, a set of values for the shortened symbols must be communicated. As we show in the results section, a protocol with a very high average efficiency can be obtained using short codes and using only three iterations.

\subsection{Blind protocol}

We formally describe below the method for blind reconciliation outlined above. Note how there is no need of an a priori error estimate (except for the one implicitly embodied in the selection of the code $\mathcal{C}$) and a controlled amount of interactivity ($t$ messages are exchanged at most).

\textit{Step 0: Set up.---} Let $\mathcal{C}(n,k)$ be a code $\mathcal{C}$ that can correct noise up to $\epsilon_{\textrm{max}}$ for some channel family. Let $X$ and $Y$ be two strings that two parties Alice and Bob wish to reconcile in at most $t$ iterations. Let $X$ and $Y$ be of length $m$, with $m=n-d$, and every symbol of $X$ and $Y$ the input and output of a memoryless channel characterized by $\epsilon < \epsilon_{\textrm{max}}$. Alice and Bob set $s=0$, $p=d$ and $\Delta=d/t$. For simplicity in the description we assume $\Delta \in \mathbf{N}$.

\textit{Step 1: Encoding.---} Alice sends the syndrome in $\mathcal{C}$ of a word $\widehat{X}$ consisting on embedding $X$ in a length $n$ string and filling the remaining $d$ positions with random symbols.

\textit{Step 2: Decoding.---} Bob constructs the word $\widehat{Y}$ consisting on the concatenation of $Y$ the received $s$ symbols and $p$ random symbols. If Bob recovers $X$ he reports success and the protocol ends.

\textit{Step 3: Re-transmission.---} If $d=s$ the protocol fails, else Alice sets $s = s + \Delta$, reveals Bob $\Delta$ symbols and they return to Step~2 and perform a new iteration.

\subsection{Average Efficiency}

The average rate of the blind reconciliation protocol can be calculated as:

\begin{equation}
\hat{R} = \sum_{i=1}^{n} \alpha_i r_i
\end{equation}

\noindent where $\alpha_i$ is the fraction, normalised to 1, of codewords that have been corrected in the iteration $i$. $r_i$ is the information rate in the same iteration. Using Eq.~(\ref{eq:efficiency-bsc}) we obtain the expression for the average efficiency over the BSC($\epsilon$):

\begin{equation}
\hat{f}_{\mathrm{BSC}}(\epsilon) = \frac{1 - \hat{R}}{h(\epsilon)} = \frac{1 - \sum_{i=1}^{n} \alpha_i r_i}{h(\epsilon)}
\label{eq:average-efficiency-bsc}
\end{equation}


Let $F^{(i)}$ be the FER when correcting with adapted rate $r_i$. Then the fraction of corrected codewords during the $i$-iteration is given by:

\begin{equation}
\alpha_i = \frac{F^{(i-1)} - F^{(i)}}{1-F^{(n)}}
\end{equation}

\noindent where $F^{(0)} = 1$.

Now, the average rate can be expressed as:

\begin{equation}
\hat{R} = \sum_{i=1}^{n} \frac{F^{(i-1)} - F^{(i)}}{1-F^{(n)}} \cdot r_i = \frac{r_1 - F^{(n)} r_n}{1-F^{(n)}} + \sum_{i=1}^{n-1} \frac{F^{(i)}}{1-F^{(n)}} (r_{i+1} - r_i)
\end{equation}

\noindent where the second equality holds for $n \geq 2$. Assuming that in every iteration we translate a constant proportion of punctured symbols to shortened symbols, the information rate used during the $i$-iteration is given by:

\begin{equation}
r_i = \frac{R_0 - \sigma_i}{1 - \delta}
\end{equation}

\noindent where $\sigma_i$ is the fraction of shortened symbols during the iteration $i$, such that $\sigma_1 = 0$ and $\sigma_n = \delta$. The rate increment between two consecutive iterations is also constant:

\begin{equation}
r_{i+1} - r_i = \frac{-\delta / (n-1) }{1 - \delta}
\end{equation}

Let us define $\beta = \delta / (1 - \delta)$ and then $r_{i+1} - r_i = -\beta / (n-1)$. The average rate can be now written as:

\begin{equation}
\hat{R} = \frac{r_1 - F^{(n)} r_n}{1 - F^{(n)}} - \frac{\beta}{n-1} \sum_{i=1}^{n-1} \frac{F^{(i)}}{1 - F^{(n)}} = r_1 + \frac{\beta}{1 - F^{(n)}} \left( F^{(n)} - \frac{1}{n-1} \sum_{i=1}^{n-1} F^{(i)} \right)
\end{equation}

Where we have taken into account that in the first iteration every selected symbol is punctured (hence $r_1= R_{max}$), while in the last one every selected symbol is shortened (hence $r_n = R_{min}$). The first and last coding rate, $r_1$ and $r_n$, are then given by:

\begin{equation}
R_{max} \equiv r_1 = \frac{R_0}{1 - \delta}
\,;\quad
R_{min} \equiv r_n = \frac{R_0 - \delta}{1 - \delta} = r_1 - \beta
\end{equation}

Note that in the rate-adaptive approach a typical value for the frame error rate in a reliable reconciliation is $10^{-3}$; i.e. we can then neglect the last contribution for the FER ($F^{(n)} \approx 0$), and thus the average rate is given by:

\begin{equation}
\hat{R} \approx r_1 - \frac{\beta}{n-1} \sum_{i=1}^{n-1} F^{(i)}
\label{eq:approx-average-rate}
\end{equation}

We describe a rough estimate of the frame error rate, $F{(i)}$, of linear codes in Appendix~A.  This analytical technique, known as the Gaussian approximation, captures the behavior of LDPC codes in the region where it transitions from correcting almost everything to correcting almost nothing (the waterfall region). Since the Gaussian approximation takes only into account the threshold and code length,  it cannot be expected to reproduce the numerical values with the same accuracy than a numerical simulation with a big sample~\cite{Yazdani_09}. It is, however, much faster. Using this approach we approximate the behaviour of a finite-length LDPC code without having to perform heavy computer simulations. 

\begin{figure} [htbp]
\centering
\epsfig{file=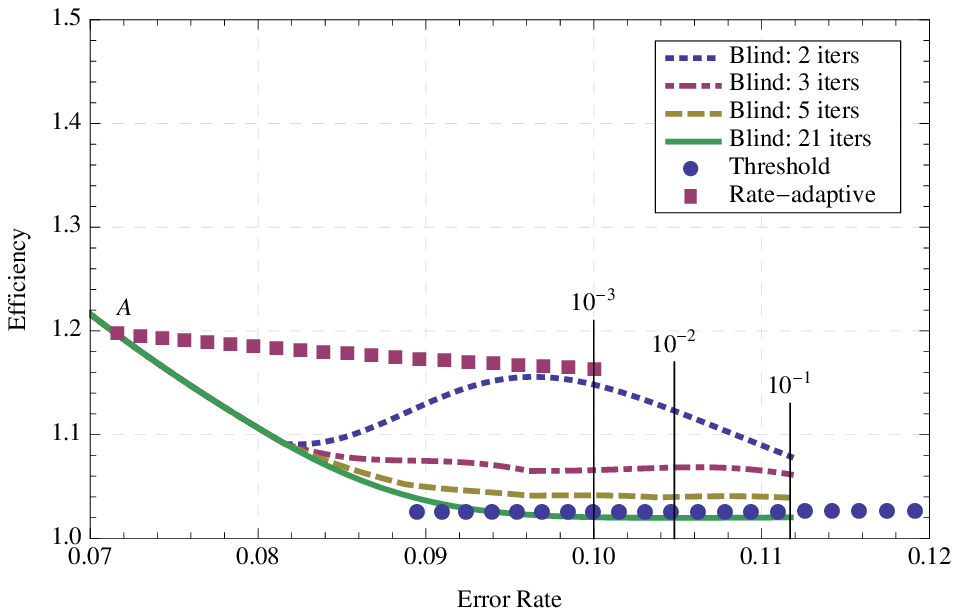, width=0.7\textwidth} \\
\epsfig{file=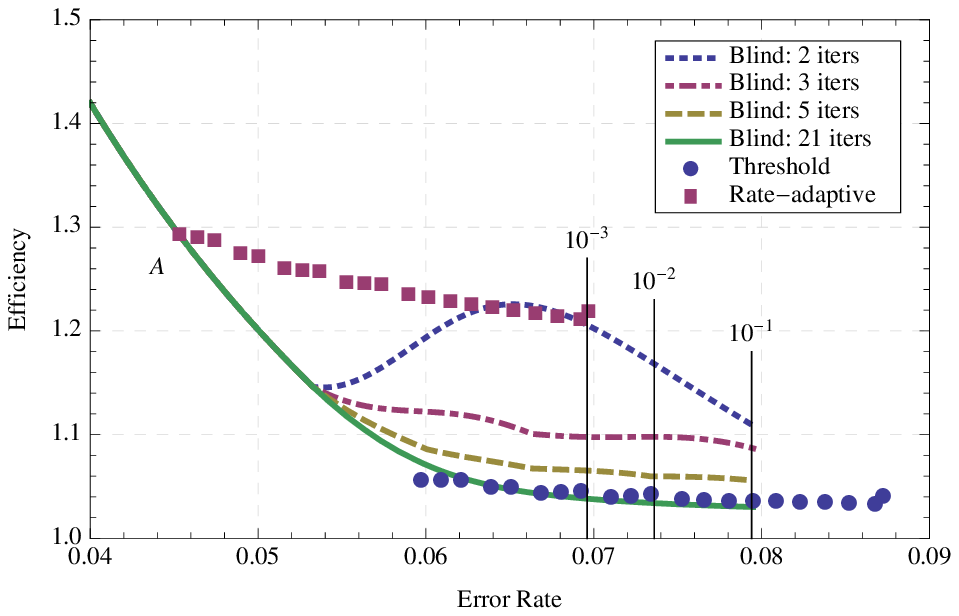, width=0.7\textwidth} \\
\epsfig{file=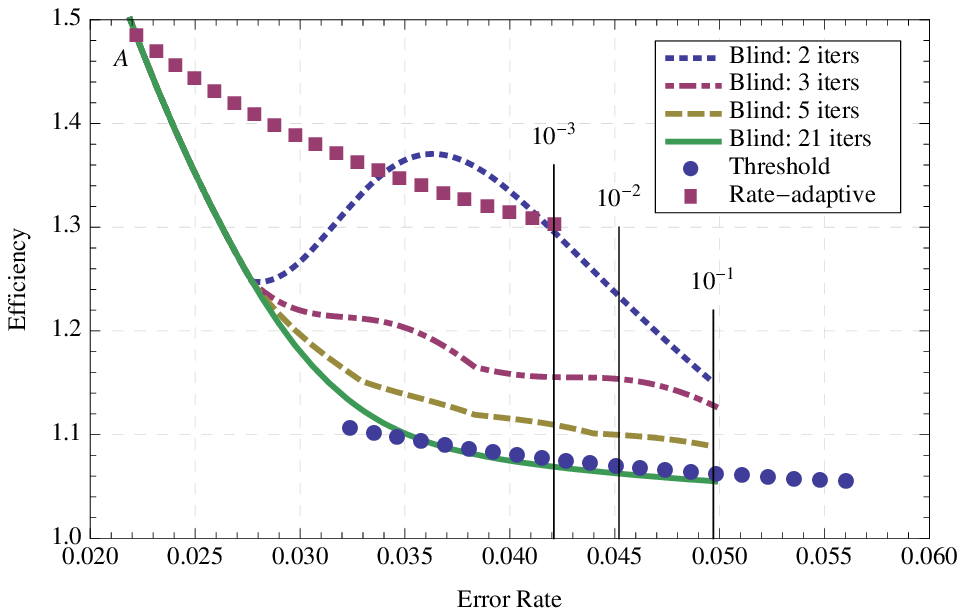, width=0.7\textwidth}
\vspace*{13pt}
\fcaption{The estimated efficiency  (Gaussian approximation) of the rate adaptive protocol is compared to the efficiency of the blind protocol for short codes ($2 \times 10^3$ bits length). Three coding rates are used to cover the error range $[2.5\%,11\%]$. A detailed description is given in the text.}
\label{fig:theoretical-efficiency-1}
\end{figure}

In Fig.~\ref{fig:theoretical-efficiency-1} we compare the decoding threshold (bullets), the rate-adaptive non-iterative protocol (boxes) and the average efficiency of the blind protocol for a different number of iterations calculated using Eqs.~(\ref{eq:average-efficiency-bsc}) and~(\ref{eq:approx-average-rate}). The figure shows the estimated efficiency for three different short-length LDPC codes of $2 \times 10^3$ bits in the error rate range $\epsilon \in [0.02, 0.11]$ with the proportion of punctured and shortened symbols set to $\delta=10\%$. From top to bottom, the coding rates are $R=0,5$, $R=0,6$ and $R=0,7$. 

Since the Gaussian approximation is a function of the decoding threshold, the maximum granularity is limited by the number of computed thresholds. We computed 21 decoding thresholds using the density evolution algorithm described in Ref.~\cite{Richardson_01a}. We stopped at 21 iterations because the curve had already converged.

The figure shows that when the error rate increases, the blind protocol adapts its behavior to the channel parameter, $\epsilon$. However, the efficiency deteriorates for low error rates. More markedly for higher coding rates.

In the figure we can differentiate three regions. First, when every symbol is punctured the redundancy is fixed, in consequence the efficiency of both, the blind protocol and the rate adaptive, coincide up to the point marked $A$. Second, the main region covers the central area: from point $A$ in the figure to the right. In this region, the blind protocol shows a better efficiency than the rate-adaptive protocol even for a low number of iterations. And third, the end of the correctable region, that depends on the maximum acceptable FER, as marked in the figure with a vertical line. In this point every symbol is shortened, i.e. this is the code with the lowest information rate.

Depending on the number of iterations allowed, which limits its maximum interactivity, the figure shows that the protocol can approach the decoding threshold. The best reconciliation efficiency for the blind protocol is achieved when it runs with $\Delta = 1$, i.e. in every iteration only one punctured symbol is converted to a shortened one. This intuitively holds because every extra intermediate code has a non zero probability of correcting which in turn increases the average coding rate. For instance, when using a code of length $2\times 10^3$ and $\delta = 10\%$ this maximum number of iterations would be 200. 

\begin{figure} [htbp]
\centering
\epsfig{file=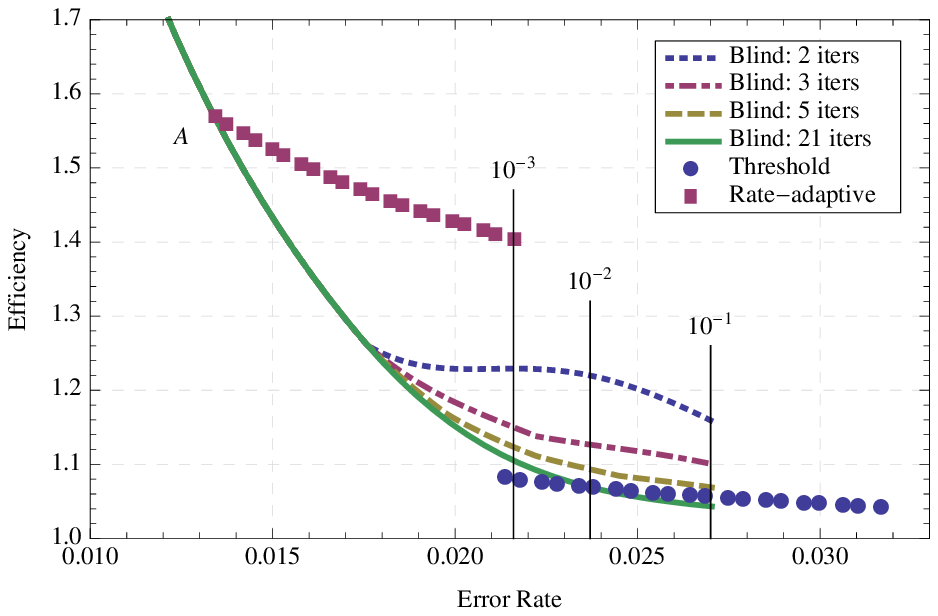, width=0.7\textwidth} \\
\epsfig{file=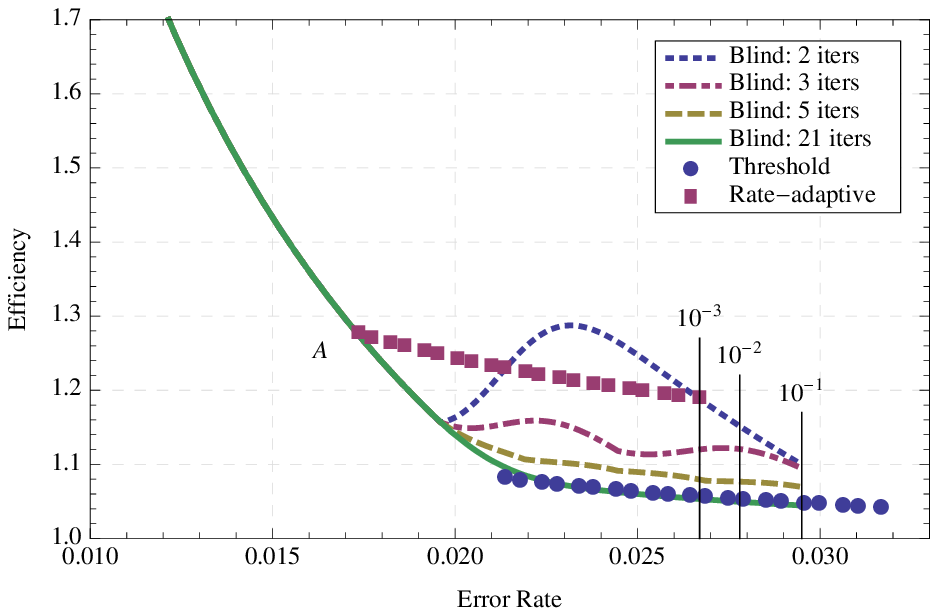, width=0.7\textwidth}
\vspace*{13pt}
\fcaption{Estimated efficiency curves, as in Fig.~\ref{fig:theoretical-efficiency-1}, for a coding rate $R=0,8$ and different codeword length, $2 \times 10^3$ (top) and $10^4$ (down) bits length. A detailed description is given in text.}
\label{fig:theoretical-efficiency-2}
\end{figure}

We cover in Fig.~\ref{fig:theoretical-efficiency-2} the lower part of the error rates of interest in QKD. In this figure we compare the performance of two LDPC codes with coding rate $R=0.8$ but different lengths ($2 \times 10^3$ (top) and $10^4$ (down) bits). The larger LDPC code was used to palliate the loss of efficiency. Codes of  length $10^4$ bits can still be implemented in hardware, although with a small throughput penalty~\cite{Burg_11}. Note that the value that $\delta$ takes has an impact in the achievable efficiency as shown in Ref.~\cite{Elkouss_11}. 
In consequence, we additionally reduce $\delta$ from $10\%$ to $5\%$ to improve the efficiency with high coding rates, thereby reducing the error rate range covered.

\section{Simulation Results}
\label{sec:results}

Simulation results were computed to compare the protocol proposed in Ref.~\cite{Elkouss_11} with the blind protocol, but using short-length LDPC codes. These simulations were performed for several error rate ranges, covering both low and high values, over the binary symmetric channel (BSC). Each simulated point was iterated till the FER was stable within a fraction of its average value. This termination condition was usually achieved when decoding between $10^2/\mathrm{FER}$ and $10^3/\mathrm{FER}$. Hence each point is the result of a very time consuming procedure that decodes between $10^5$ and $10^8$ codewords, depending on the FER. An LDPC decoder based on the sum-product algorithm was used. New LDPC codes were designed for several coding rates (see Appendix~B) and constructed using Ref.~\cite{Kim_07}. This construction focus in reducing the error floor\footnote{The residual error in the low error rate region.}. The objective of considering low error floor construction was to improve the efficiency of the blind protocol as it corrects words in a wide error rate region.

The shortened symbols were selected randomly while the punctured symbols were selected according to a computed pattern for intentional puncturing as described in Ref.~\cite{Elkouss_11b}. This intentional puncturing algorithm is specifically designed for moderate puncturing rates, i.e. low values of $\delta$.

\begin{figure} [htbp]
\centering
\begin{tabular}{c}
\centerline{\epsfig{file=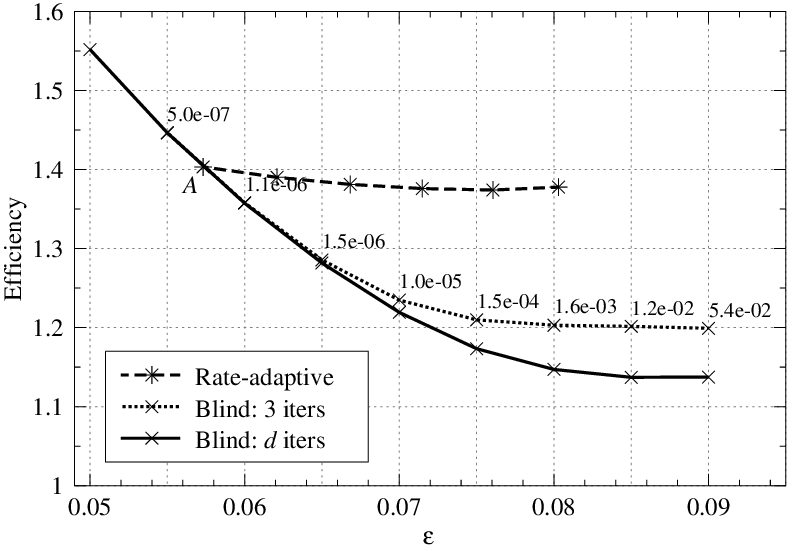,width=0.7\textwidth}} \\
\centerline{\epsfig{file=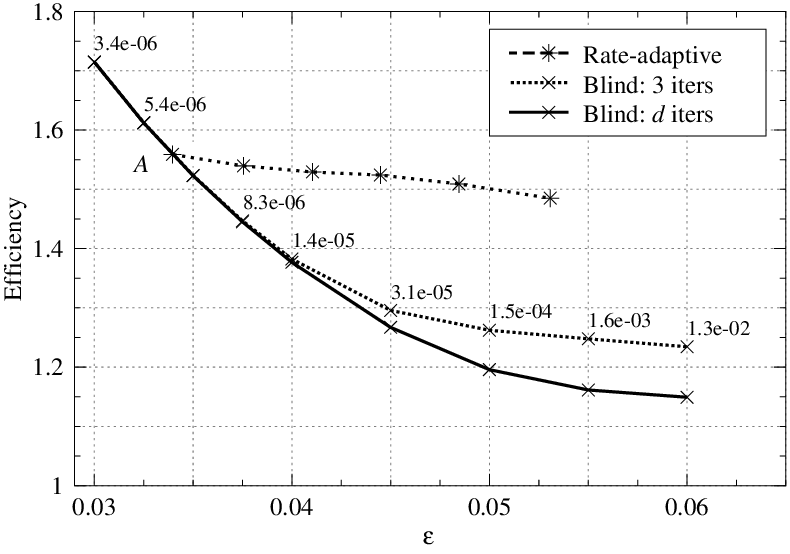,width=0.7\textwidth}}
\end{tabular}
\fcaption{Simulated curves for the rate-adaptive protocol proposed in Ref.~\cite{Elkouss_11} and the blind protocol. LDPC codes of $2 \times 10^{3}$ bits length and rates $R=0.5$ (top) and $R=0.6$ (bottom) were used. Punctured symbols were selected according to a pattern for intentional puncturing as proposed in Ref.~\cite{Elkouss_11b}. The average FER is printed for each point of the three iterations curve. The lowest curve is for the $d=n*\delta=200$ iterations. Every point for the rate-adaptive protocol was determined for a frame error rate of $10^{-3}$. }
\label{fig:simulated-efficiency-1}
\end{figure}

Fig.~\ref{fig:simulated-efficiency-1} shows the efficiency, as defined in Eq.~(\ref{eq:efficiency}), of the rate-adaptive protocol proposed in Ref.~\cite{Elkouss_11}, and the average efficiency of the blind protocol for two different rates. The blind protocol is simulated with two different maximum number of iterations or equivalently with two different values of $\Delta$. The first one, a lightweight version limited to a maximum of 3 iterations, is compared with the maximally interactive version where in every iteration only one punctured symbol becomes a shortened one ($\Delta = 1$). 
Simulations were computed using an LDPC code of $2 \times 10^{3}$ bits length with $\delta=10\%$ and two coding rates $R=0.5$ and $R=0.6$.

The simulations for the rate-adaptive approach were computed with an acceptable FER set to $10^{-3}$. However, in this figure and in the following one, the curves for the blind protocol extend beyond the acceptable FER. To show its value, in the version with a maximum of three iterations the average FER is printed.

The figure shows how the efficiency improves with interactivity (more iterations) and with the error rate.  The efficiency in Fig.~\ref{fig:simulated-efficiency-1} also coincides for the rate-adaptive and interactive approaches for error rates below $A$, a behaviour similar to Fig.~\ref{fig:theoretical-efficiency-1}.

\begin{figure} [htbp]
\vspace*{13pt}
\begin{tabular}{c}
\centerline{\epsfig{file=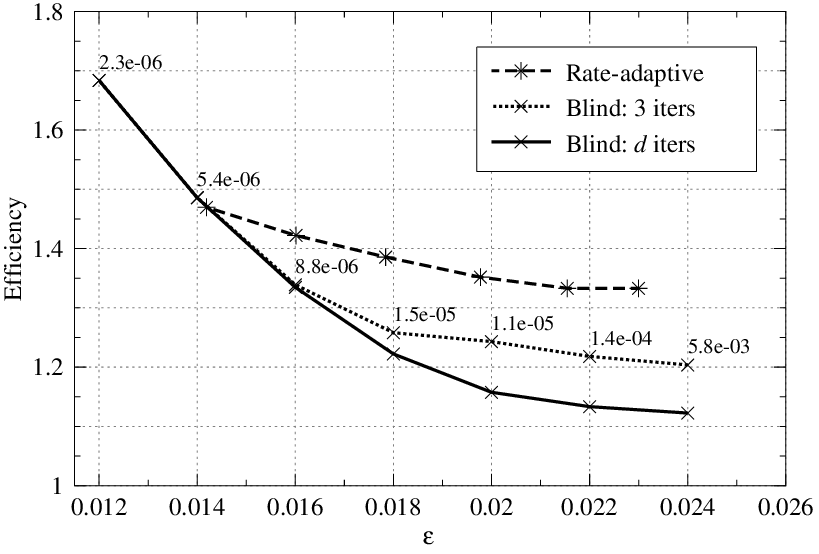, width=0.7\textwidth}} \\
\centerline{\epsfig{file=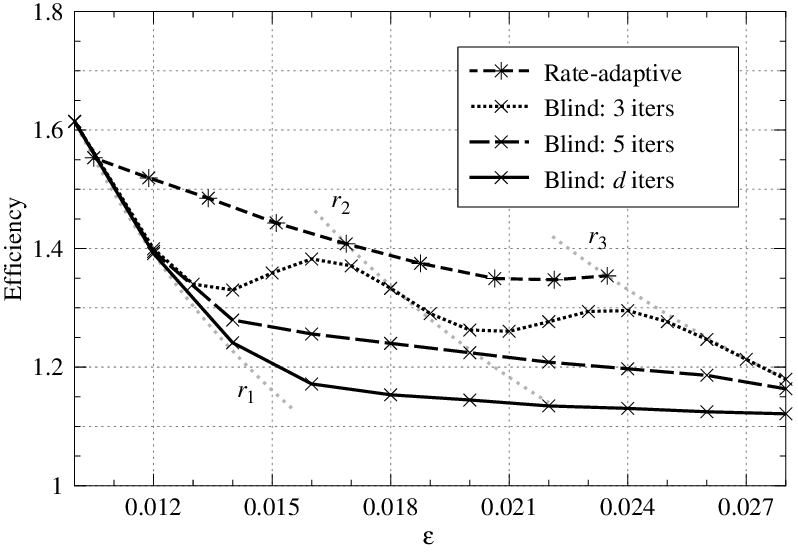, width=0.7\textwidth}} 
\vspace*{13pt}
\end{tabular}
\fcaption{Simulated curves for the rate-adaptive protocol proposed in Ref.~\cite{Elkouss_11} and the blind protocol. An LDPC code of $10^{4}$ bits length and rate 0.8 was used. Symbols were selected according to an improved pattern for intentional puncturing as proposed in Ref.~\cite{Elkouss_11b}. The average FER is printed for each point of the three iterations curve (top). The lowest curve is for $d=n*\delta=1000$ iterations. Every point for the rate-adaptive protocol was determined for a frame error rate of $10^{-3}$. 
In order to understand the behaviour of the curve for the blind protocol with a maximum of 3 iterations, we also plot (bottom) the efficiency of using (fixed-rate) LDPC codes with the coding rates associated with each iteration: $r_1 = \frac{R_0}{1 - \delta}$, $r_2 = \frac{R_0 - \delta/2}{1 - \delta}$ and $r_3 = \frac{R_0 - \delta}{1 - \delta}$.}
\label{fig:simulated-efficiency-3}
\end{figure}

In Fig.~\ref{fig:simulated-efficiency-3} the efficiency is studied in the low error rate range in QKD. An LDPC code of $10^{4}$ bits length and coding rate $R=0.8$ is used. Due to this high coding rate, only $5\%$ of the symbols were selected for puncturing and shortening. The figure shows that the average efficiency quickly improves with the blind protocol, even when using only three iterations.

If we try to increase the range of error rates covered, we can increase the proportion of punctured and shortened symbols (see Eq.~(\ref{eq:rate-range})). The results are shown in the bottom panel of Fig.~\ref{fig:simulated-efficiency-3}, where the proportion is set to the same $8\%$, the maximum achievable value following the intentional puncturing proposal described in Ref.~\cite{Elkouss_11b}. We observe that for a fixed number of iterations the efficiency is worse, as is clearly seen when comparing the dotted line (with a maximum of three iterations) in both panels. As expected, the efficiency for the maximum number of iterations, $d$, as it grows, improves.

The increase in efficiency with the number of iterations opens the possibility of having both, high efficiency and high throughput. The new generation of QKD systems are approaching sifted-key rates close to $1$ Mbps \cite{Takesue_05, Tang_06, Shields_09, Choi_10, Zbinden_10, Buller_10}. Implementing real time error correction to provide secret keys at this speed is a challenging problem where a high throughput procedure with minimal communications is needed. Using \textit{Cascade} under these constraints is unfeasible unless an extremely low latency network is used. Short-length LDPC codes have been implemented in hardware for other purposes, like wireless networks \cite{IEEE_802.11n_09}, where they have demonstrated to be an excellent solution. In the QKD case, to use LDPC codes required to have codes designed for different error rates, thus making the process more complex and memory constrained. With the protocol presented here, the error estimation phase is not needed. The procedure can start directly and, if it fails, allowing a few iterations increases considerably the success probability. The price to pay is an extra message per failure. As shown in Figs.~\ref{fig:theoretical-efficiency-1} and~\ref{fig:simulated-efficiency-3}, the process converges quickly and only a few iterations are needed to increase the reconciliation efficiency significantly. This also avoids the need to store many precalculated codes.

In a hardware implementation, the iterations are easily realised just by copying the same functional decoder block as many times as the number of desired iterations. The string to be reconciled would start the iteration $i$ in the first hardware block. If the decoding fails, the next hardware block would continue processing in a pipeline fashion, since computation and communication can be arranged in a way such that the disclosed symbols would arrive packed in the same message than the syndrome of the following strings to reconcile. The new syndrome would start being processed in the first hardware block while the second would continue working on the second iteration on the previous string. This pipeline can increase the efficiency while maintaining a high and ---mostly--- constant throughput at the expense of some extra hardware.

\section{Conclusions}
\label{sec:conclusions}

Results show that efficient reconciliation can be achieved using short-length LDPC codes and a slightly interactive protocol. Codes as short as 2000 bits are suitable for blind reconciliation. Even a minimum interactive protocol with a maximum of three iterations improves considerably the efficiency in high and low error rate regimes. 

This protocol can be easily implemented in hardware and it can be also pipelined to increase the throughput of reconciled key. Its requirements are low enough to allow an implementation using an embedded processor within a compact, industrial grade QKD device.

The protocol presented here could improve the final secret-key length in those scenarios where part of the raw key has to be disclosed in order to estimate the channel parameter. This improvement lies in the fact that the protocol works without an error rate estimate.

\nonumsection{Acknowledgements}

The authors wish to thank Andreas Burg for helpful discussions about the hardware implementation of LDPC codes. The authors also gratefully acknowledge the computer resources, technical expertise and assistance provided by the \emph{Centro de Supercomputaci\'on y Visualizaci\'on de Madrid}\footnote{http://www.cesvima.upm.es} (CeSViMa) and the Spanish Supercomputing Network.

This work has been partially supported by the project Quantum Information Technologies in Madrid\footnote{http://www.quitemad.org} (QUITEMAD), Project P2009/ESP-1594, \textit{Comunidad Aut\'onoma de Madrid}.

\nonumsection{References}

\appendix{~Frame Error Rate Analysis}

In this work we study the FER for finite-length communications using the concept of \emph{observed} channel introduced in Ref.~\cite{Yazdani_09}. Though a simple approach, it provides with an acceptable accuracy. The method is based on the analysis of the probability density function (pdf) of random variables corresponding to $N$ received symbols. This pdf for a $N$ finite-length code is compared with its estimated threshold, $\varepsilon^*$, such that an average frame error rate probability is calculated as the probability that the ``observed'' channel behaves worse than the code's decoding threshold. This observed channel is interpreted as the measurement of $N$ samples, where each sample (bit) is an error with probability $p$.

\textit{Observed Channel.---} Any communication channel (discrete memoryless channel) is stochastically modelled by a set of parameters. For instance, the binary symmetric channel (BSC) is parameterised by its error rate $\epsilon$. However, these parameters accurately describe the behaviour of the modelled channel only in the asymptotic case, i.e. assuming infinite length communications. In the BSC($\epsilon$) we define the \textit{observed} bit error rate in a communication, $P_{\mathrm{obs}}$, as the number of errors divided by the length of this communication, $N$. This observed value is constant only in the asymptotic case, i.e. $P_{\mathrm{obs}} = \epsilon \; \forall N$ only when $N \to \infty$. The distribution of errors in our \textit{observed} BSC channel is then described by the following probability mass function (pmf):

\begin{equation}
f_{P_{\mathrm{obs}}}(\epsilon, N, x) = \binom{N}{Nx} \epsilon^{Nx} (1-\epsilon)^{N-Nx}
\end{equation}

\noindent where $Nx$ is the number of errors in the communication of length $N$.

Assuming that the length of the communication is large enough (i.e. when it is higher than a few thousand bits) this pmf can be approximated with high precision by using a (continuous) Gaussian probability density function centered around the error rate $\epsilon$ and with variance $\sigma_{P_{\mathrm{obs}}}^2 = \epsilon (1-\epsilon) / N$:

\begin{equation}
f_{P_{\mathrm{obs}}}(\epsilon, N, x) \approx \mathcal{N}(\mu_{P_{\mathrm{obs}}}, \sigma_{P_{\mathrm{obs}}}^2)
\end{equation}

\textit{Frame Error Rate.---} Let us now consider that we are using a finite-length linear code to correct any error occurred during the communication, then we can estimate the ratio of codewords that cannot be corrected by calculating the probability that the observed channel behaves worse than the decoding threshold of our code, $\epsilon^*$ (see Ref.~\cite{Richardson_01a}).

Using an error correction code of length $N$ with a theoretical threshold of $\epsilon^*$, the FER for our BSC($\epsilon$) channel can be reasonably approximated by:

\begin{eqnarray}
F_{P_{\mathrm{obs}}}(\epsilon, N, \epsilon^*) & = & 1 - \Pr(P_{\mathrm{obs}} \le \epsilon^*) \\
& = & \Pr(P_{\mathrm{obs}} > \epsilon^*) = \int_{\epsilon^*}^1 f_{P_{\mathrm{obs}}} (\epsilon, N, x) dx
\label{eq:fer-observed-channel}
\end{eqnarray}

Using the Gaussian approximation:

\begin{equation}
F_{P_{\mathrm{obs}}}(\epsilon, N, \epsilon^*) \approx \frac{1}{\sqrt{2\pi \epsilon (1-\epsilon) / N}} \int_{\epsilon^*}^1 e^{-\frac{N (x-\epsilon)^2}{2\epsilon (1-\epsilon)}} dx
\end{equation}

Note that, for convenience, we have used the term $F$ instead of $F_{P_{\mathrm{obs}}}(\mathcal{C}(\theta), N, \epsilon^*)$ in the main body of the paper.

Note also that this analytical approximation is only valid for the behaviour in the \emph{waterfall} region of an error correction code, since it does not include information about the performance in the error floor regime.

\appendix{~Families of LDPC Codes}

In this appendix we present the generating polynomials (see Ref.~\cite{Richardson_01a}) for the LDPC codes used in this paper. The design criteria was to maximise the threshold for a ratio of edges/bit not greater than $6.06$. A small ratio of edges/bit reduces the achievable threshold but renders the codes more suitable for hardware implementations.

Coding rate $R = 0.5$, theoretical threshold $\epsilon^* = 0.102592$ (see Ref.~\cite{Richardson_01a}):

\begin{eqnarray}
\lambda(x) & = & 0.159673 x + 0.121875 x^{2} + 0.11261 x^{3} + 0.190871 x^{4} + \nonumber \\
 & & 0.0770616 x^{9} + 0.337909 x^{24} \nonumber \\
\rho(x) & = & 0.360479 x^{8} + 0.639521 x^{9}
\end{eqnarray}

Coding rate $R = 0.6$, theoretical threshold $\epsilon^* = 0.0745261$:

\begin{eqnarray}
\lambda(x) & = & 0.11653 x + 0.125646 x^{2} + 0.108507 x^{3} + 0.0534223 x^{4} + \nonumber \\
 & & 0.0727228 x^{6} + 0.0347964 x^{7} + 0.0729986 x^{8} + \nonumber \\
 & & 0.0752607 x^{17} + 0.117103 x^{31} + 0.223013 x^{44} \nonumber \\
\rho(x) & = & 0.582731 x^{13} + 0.417269 x^{14}
\end{eqnarray}

Coding rate $R = 0.7$, theoretical threshold $\epsilon^* = 0.0501875$:

\begin{eqnarray}
\lambda(x) & = & 0.091699 x + 0.171401 x^{2} + 0.0683878 x^{3} + 0.120523 x^{4} + \nonumber \\
 & & 0.187471 x^{10} + 0.208278 x^{27} + 0.152239 x^{29} \nonumber \\
\rho(x) & = & 0.806453 x^{18} + 0.193547 x^{19}
\end{eqnarray}

Coding rate $R = 0.8$, theoretical threshold $\epsilon^* = 0.0289413$:

\begin{eqnarray}
\lambda(x) & = & 0.0667948 x + 0.194832 x^{2} + 0.0570523 x^{3} + 0.0645024 x^{4} + \nonumber \\
 & & 0.204606 x^{8} + 0.0964409 x^{14} + 0.23872 x^{28} + 0.0770523 x^{34} \nonumber \\
\rho(x) & = & 0.708874 x^{29} + 0.291126 x^{30}
\end{eqnarray}

\end{document}